\documentclass{emulateapj}
\usepackage{graphicx}
\usepackage{epsfig}

\begin{document}

\title{The essence of quintessence and the cost of compression}
\author{Bruce A. Bassett}
\affil{Department of Physics, Kyoto University, Kyoto, Japan}
\affil{Institute of Cosmology and Gravitation, University of 
Portsmouth, 
Portsmouth, PO1 2EG}
\author{Pier Stefano Corasaniti}
\affil{ISCAP, Columbia University, New York, NY, 10027, USA}
\and
\author{Martin Kunz}
\affil{Astronomy Centre, University of Sussex, Brighton, BN1 9QJ, UK}
\begin{abstract}
Standard two-parameter compressions of the infinite dimensional dark energy model space show 
crippling limitations even with current SN-Ia data. Firstly they cannot cope with 
rapid evolution - our best-fit to the latest SN-Ia data shows late and 
very rapid evolution to $w_0 = -2.85$. However all of the standard parametrisations 
(incorrectly) claim that this best-fit is ruled out at more than $2\sigma$ primarily
because they track it well only at very low redshift, $z \leq 0.2$. Further they incorrectly rule 
out the observationally compatible region $w \ll -1$ for $z>1$. Secondly the parametrisations give wildly different estimates for the redshift of acceleration, which vary from $z_{acc}=0.14$ to $z_{acc}=0.59$. 
Although these failings are largely cured 
by including higher-order terms ($\geq 3$ parameters) this results in new degeneracies 
and opens up large regions of previously ruled-out parameter space. Finally we test 
the parametrisations against a suite of theoretical quintessence models. The widely used linear expansion
in $z$ is generally the worst, with errors of up to $10\%$ at $z=1$  and $20\%$ at $z \geq 2$. 
All of this casts serious doubt on the usefulness of the standard two-parameter 
compressions in the coming era of high-precision dark energy cosmology and  
emphasises the need for decorrelated compressions with at least three parameters. 
\end{abstract}


\section{Introduction}
The issue of dark energy dynamics is perhaps the most pressing today in 
cosmology. There are claims both for and against dynamics (Bassett {\em 
et al.} 2002, Alam {\em et al.} 2003, Jassal {\em et al.} 2004). But it 
is a subject dogged by gauge problems \footnote{Consider two dark 
energy parametrisations, $P_1$ and $P_2$. We are interested in estimating 
the true value of an observable ${\cal O}$, such as $w(z=0)$. Under a 
change in parametrisation the best estimate of ${\cal O}$ will change to 
${\cal O}_2 = {\cal O}_1 + \delta {\cal O}$. This is a gauge artifact 
since the change in ${\cal O}$ has nothing to do with the real 
universe. Even worse, since the two parametrisations will usually be rather 
different we have no guarantee that $\delta {\cal O}$ will be small.} 
\cite{Jonsson,Virey,WT}. 

Claims for dynamics (or the lack thereof) can be pure gauge artifacts, 
mirages induced by the parametrisations that are unrelated to the data. For 
example,  Riess {\em et al.} (2004) and Jassal {\em et al.} (2004) 
claim that current SN-Ia data are inconsistent with rapid evolution of dark 
energy. Such conclusions must always implicitely refer to a finite 
dimensional subspace of the full dark energy model space, and broadening 
the class of models studied can (and in this case does) lead to complete 
reversal of such conclusions. Fig. (\ref{fig6}) provides a explicit
counter-example.  

The aim of compression is to summarise, as aggressively as possible, 
the key features of dark energy 
properties in a few parameters and to facilitate discrimination between 
models. We will see that standard expansions fail to achieve either of 
these goals, even when they are extended to higher order. 

The main result of this work is that compression of the dark energy 
space into low-dimensional subspaces, while convenient and easy to work 
with, can give seriously misleading results. Since these results are used 
in the design of upcoming surveys it is bad news for cosmology in 
general. If one does not impose the weak energy condition (WEC), $w \geq 
-1$, then the results can border on the completely useless. The rest of 
this article delimits, as precisely as possible, the quicksands and danger areas
in the use of two-parameter compressions.

As a first sobering example, consider constraints on $w(z)$ when we do 
not impose the WEC. One-parameter studies give constraints such as 
$-1.38< w_{DE} <-0.82$ at $2\sigma$ \cite{alm} suggesting that a model with 
$w=-5$ at $z=2$ would be ruled out at more than $10\sigma$. 
Instead a little thought  makes it clear that if $w(z)$ can vary freely 
then  {\em there is no lower 
bound on $w$} for $z \geq 1$ since this merely changes how fast the 
already irrelevant and rapidly diminishing dark energy density decreases. 
If the rapid drop in $w$ occurs at $z>1$ this leaves essentially no 
observable trace \cite{Bassett,Coras3}. This is clearly reflected in the 
likelihoods in Fig. (\ref{fig2}) which allow for $w < -100$ at $z\sim 1$.  How 
can we hope to cover such possibilities with simple one or two parameter 
compressions? 
 
The dark energy literature overflows with one, two and higher-dimensional compressions of 
$w_{DE}(z)$, e.g. (Efstathiou 1999; Huterer and Turner 2001; Weller and Albrecht 
2002; Bassett et al. 2002; Corasaniti and Copeland 2002; Linder 2003; Jassal 
{\em et al.} 2004). Compressions also exist for $\rho_{DE}(z)$ \citep{Wang,Wetterich} while 
decorrelated reconstructions of $w_{DE}(z)$ have been proposed in \citep{Huterer02} and \citep{hu04}. 

The most basic parametrisation, namely describing the dark energy with a constant equation 
of state $w=w_0$, is well-known to introduce a severe bias (see for instance Maor {\em et al.}
2002, Virey {\em et al.} 2004) in parameter estimation. 
Compressions invoking two parameters which somewhat 
alleviate this problem have been introduced in 
\citep{Efstathiou99,Linder03,Jassal}. However, as we will see, these models all 
struggle to describe rapid evolution. This is not surprising. With two 
parameters one may fix $w$ at $z=0$ and $w$ at high $z$, but one can do 
nothing about the time nor the rapidity of the transition between the two extremes.
Caldwell and Doran (2004) circumvented this by considering thirteen different one and 
two-parameter models, some exhibiting rapid transitions.  

The use of more than two parameters offers the opportunity to test 
the dark energy evolution with different data sets and consistently 
account for the dark energy perturbations at {\em all} redshifts  (Bassett {\em et al}. 
2002; Corasaniti and Copeland 2002; Corasaniti {\em et al.} 2004) but is, 
of course, more computationally intensive, and care must be taken to accurately 
capture the dark energy dynamics of rapid transitions; see Appendix C 
of (Corasaniti {\em et al.} 2004).

\section{The parametrisations}\label{params}

For our study we consider two distinct classes of compressions. First are standard 
Taylor expansions of $w_{DE}(z)$ and second is the {\em Kink}, a 
physically-motivated compression. The Taylor expansions are all of the form:
\begin{equation} 
w_{DE}(z)= \sum_{n=0} w_n x_n(z)
\label{taylor}
\end{equation}
where we consider four different choices for the `expansion' functions, 
$x_n(z)$. Namely:
\begin{eqnarray}
x_0(z)&&\equiv 1 \,; \, x_n \equiv  0,~ \,n \geq 1 ~~~~~~~~~( {\rm \bf 
constant~} w)\\
x_n(z)&& \equiv z^n \,\,~~~~~~~~~~~~~~~~~~~~~~~~~~~~({\rm \bf 
redshift}) \label{lin}\\
x_n(z)&& \equiv (1-a)^n =\left(\frac{z}{1+z}\right)^n \,~~~ ({\rm \bf 
scale~factor}) \label{linder}\\
x_n(z)&& \equiv (\log(1+z))^n \, ~~~~~~~~~~~~~~~({\rm \bf logarithmic}) 
\label{log}\\
&& \nonumber
\end{eqnarray}

To linear order ($n\leq 1$) these were first discussed by Huterer and 
Turner (2001) \& Weller and Albrecht (2002), Chevallier and Polarski (2000) and
Linder (2003), and Efstathiou (1999)  for the redshift, scale-factor and 
logarithmic expansion functions respectively. 
Later we will consider their performance at higher order ($n \geq 2$).

The Kink, on the other hand, is not an expansion. It is a  4-parameter 
model which accurately captures the behaviour of quintessence  (Bassett 
{\em et al.} 2002, Corasaniti and Copeland 2002, Corasaniti {\em et 
al.} 2004). The extra parameters allow us to model very rapid transitions 
in $w_{DE}(z)$, a freedom we will need:
\begin{equation}
w_{DE}(a)=w_0+(w_m-w_0)\frac{1+e^{\frac{a_t}{\Delta}}}{1+e^{\frac{(a_t-a)}{\Delta}}}
\frac{1-e^{\frac{(1-a)}{\Delta}}}{1-e^{\frac{1}{\Delta}}},\label{bcc}
\end{equation}
$a$ is the scale factor, $w_0$ and $w_m$ are the present and matter-dominated values of 
the dark energy equation of state, $a_t$ is the value of the scale factor at 
the transition from $w_m$ to $w_0$ and $\Delta$ controls the width of the transition.
Other formulations of the Kink, with relative merits, 
are discussed in Appendix A of Corasaniti {\em et al.} (2004).

There are other parametrisations but  these are the most widely used 
today and lessons learned from these compressions will apply to many others 
in the literature. 

\section{Constraints from SN-Ia \label{sn1a}} 

In this section we investigate whether the different parametrisations above 
give rise to different best-fits to current type Ia supernova (SN-Ia) data. 
Recently $6$ new SN-Ia at redshift $z>1.25$ have been discovered using the Hubble Space Telescope,
providing further evidence for a transition from  
decelerated to accelerated expansion in the past\citep{Riess}. 

\begin{figure}[t]
\plotone{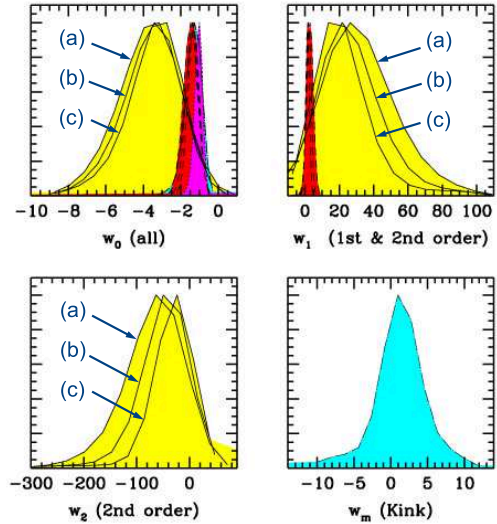}
\caption{{\bf Expansion order is more important than parametrisation}. 
1-d marginalised likelihoods for the different classes of 
parametrisations. The narrow curves are the 
linear likelihoods (constant $w_{DE}$, $n=0$, is dotted, first-order expansions, $n\leq 1$, 
are dashed). The solid lines show the likelihoods at second-order ($n\leq 2$) and 
are much wider due to degeneracies and the absence of a lower-bound on $w$ for $z\geq 1$. The 
curves correspond to expansions in (a)  scale-factor, (b) log and (c) redshift respectively. Finally the 1-d likelihood for $w_m$ of  the Kink is shown (dot-dashed) and exhibits strongly non-Gaussian wings 
that extend to very large values of $|w_m|$. For definition of the order of the expansion see eq. (\ref{taylor}).
\label{fig2}}
\end{figure}

In order to be conservative we use
only the gold sample of \citep{Riess}, containing $157$ data points.
In our analysis we assume a flat Friedmann-Lemaitre-Robertson-Walker 
(FLRW) universe. The assumption of flatness is required to achieve 
reasonable error bars (c.f. Kunz and Bassett 2004; Dicus and Repko 2004). 
Fortunately this is now a data-driven assumption and particularly 
harmless for this study since we are 
primarily interested in testing compressions rather than deriving constraints. 
We will also assume the prior $\Omega_m = 0.27 \pm 0.04$. 
This can be justified from CMB data, as shown in Kunz {\em et al.} (2003) 
and Corasaniti {\em et al.} (2004),  the best-fit values for 
background FLRW parameters are not affected 
strongly by dark energy dynamics. 

The luminosity distance is given by
\begin{equation}
d_L(z)= (1+z)\int_0^z \frac{dz'}{H(z')},
\end{equation}
and
\begin{equation}
H(z)=H_0 [\Omega_m(1+z)^3+(1-\Omega_m)f(z)]^{1/2},
\end{equation}
where $f(z)$ is the solution to the continuity equation
\begin{equation}
f(z)=\exp{\left(3\int_0^{z} \frac{1+w_{DE}(z')}{1+z'}dz'\right)}.
\end{equation}
Our analysis methods are described in detail in Corasaniti {\em et al.} (2004).
We use a Markov-Chain Monte Carlo code to find the constraints on the 
dark energy parameters for each parametrisation. As usual we marginalise 
analytically over the normalisation of $d_L$ which takes care of the 
Hubble constant as well, leaving $\Omega_m$ as the only remaining 
parameter apart from those describing the equation of
state.  

Figure \ref{fig2} shows the marginalised one-dimensional 
likelihoods for the parameters of the dark energy compressions. The main point of that 
figure is that the various parametrisations have similar likelihoods {\em at the same order}, but that 
the likelihoods at different orders are completely different. Here by order we mean the 
maximum value of $n$ in eq. (\ref{taylor}). Hence ``linear" or ``first-order" ($n\leq 1$) refers to the standard two-parameter expansions with only $w_0,w_1$ non-zero, while 2nd-order corresponds to $n\leq 2$ and has non-zero $w_2$. 

The best-fit values and $1\sigma$ errors for the first-order 
redshift, scale factor and logarithmic parametrisations are given in table \ref{tab1}; these 
are consistent with those found in (Feng et al. 2004; Gong 2004).

%%%%%%%%%%%%%%%%%
\begin{table}[ht]
\caption{Best fit values and $1\sigma$ confidence intervals at linear 
order ($n\leq 1$).\label{tab1}}
\begin{center}
\begin{tabular}{crrr}
\tableline\tableline
& $w_0$ & $w_1$ \\ 
\tableline\tableline\\
Redshift  &$-1.30\pm_{0.52}^{0.43}$&$1.57\pm_{1.41}^{1.58}$\\\\
Scale-factor &$-1.48\pm_{0.64}^{0.57}$&$3.11\pm_{3.12}^{2.98}$\\\\
Log    &$-1.39\pm_{0.57}^{0.50}$&$2.25\pm_{2.15}^{2.19}$\\\\
\tableline
\end{tabular}
\end{center}
\end{table}

\begin{figure}[ht]
\plotone{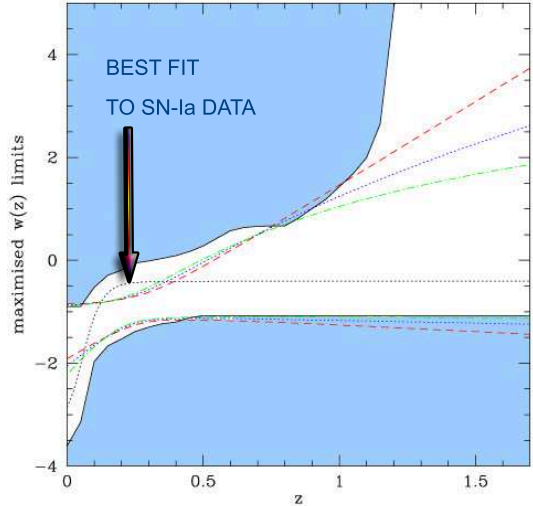}
\caption{{\bf Parametrisations struggle with rapid evolution.} Maximised limits on 
$w_{DE}(z)$  for the redshift (red dashed line),
scale-factor (green dash-dotted), logarithmic (blue dotted) and Kink (solid black line) 
parametrisations. The best-fit kink solution passes well outside the limits 
of all the parametrisations (except for the kink) both at $z\sim 0$ and at $z\sim 0.2$, 
showing their inability to  capture rapid dynamics which  
leads to their incorrectly ruling it out. \label{fig6}}
\end{figure}

We infer the 'maximised' limits on the redshift dependence of the 
equation of state by computing
for a given parametrisation the highest and lowest $w_{DE}(z)$ for the 
models in the chains
with $\chi^2<\chi_{min}^2+4$. Models with $w(z)$ outside those limits 
are therefore
expected to be ``bad'' fits to the SN-Ia data.
We plot the result in Figure~\ref{fig6}. For the two-parameter
models, the limits are very similar to those obtained by marginalising 
over all other
parameters, as is expected since they are nearly Gaussian. 
We have also checked that they
coincide with limits from Gaussian error propagation.
%%%%%%%%%%%%%%%%%%%%%%%%%%%%%%%%%%%%%%%%%%%%%%%%%%%%%%%%%%%%%%%%%%%%%%%%%%%%%%%%%%%%%%%%%%%%%55
%%%%%%%%%%%%%%%%%%%%%%%%%%%%%%%%%%%%%%%%%%%%%%%%%%%%%%%%%%%%%%%%%%%%%%%%%%%%%%%%%%%%%%%%%%%%%%%%%%

We now discuss the constraints derived from the Kink formula, 
Eq.~(\ref{bcc}).
The best fit to the data has a $\chi^2=172.6$ and
is characterized by:
$w_0=-2.85$, $w_m=-0.41$, $a_t=0.94$ and $\log({\Delta})=-1.52$,  
corresponding to a rapidly varying equation of state with a transition
from $w_m$ to $w_0$ at $z_t=0.1$. This best-fit is shown in figure \ref{fig6} and
clearly exits the $2\sigma$ limits from the two-parameter compressions, first from below at $z \sim 0$ 
and then from above, at $z\sim 0.2$. This graphically illustrates the limitations of the 
standard parametrisations and shows how they artificially rule out models which should give 
the strongest signals for dark energy dynamics \cite{Coras3}.

Although the Kink formula is by construction everywhere under control,
the dependence on its parameters is highly non-linear so the 
peaks of the marginalised likelihood distributions 
do not match the values of the  best fit model.
In fact the peaks of the $1$-dimensional marginalised likelihoods are 
shifted with the respect to the best fit values after marginalising. Hence the 
marginalised likelihoods do not coincide with the maximized ones, as they would 
if the likelihood distributions  were Gaussian.

It is not just one good fit to the data which violates the $2\sigma$ limits of all the two-parameter 
compressions either.  For instance, the model with
$w_0=-1.46$, $w_m=0.16$, $a_t=0.88$ and $\log{(\Delta)}=-0.7$ has 
$\chi^2=173.9$, while the model with $w_0=-1.11$, $w_m=6.13$, $a_t=0.40$ and 
$\log{(\Delta)}=-.98$, has $\chi^2=175.9$. Both are excellent fits to the data but 
are supposedly ruled out by the $n\leq 1$, linear redshift, scale-factor and logarithmic 
parametrisations of equations (\ref{lin}-\ref{log}). 

The conclusion that rapid evolution of dark energy is ruled out by current data is
therefore a `gauge' artifact. We have shown that rapid 
variations of the dark energy equation of state are perfectly consistent with, and in
fact provide better fits to the Gold sample than do models without rapid transitions.
This conclusion remains even after including CMB and large scale structure data (Corasaniti
{\em et al.} 2004).

The pathological behaviour of ruling out models which are very good fits to the data 
can be rectified by the inclusion of higher order terms, $n\geq 2$. 
Indeed, since the data allow $w_1$ to be large in all cases, 
higher order terms in the redshift, scale-factor
and Log expansions cannot be neglected. Therefore we have extended our 
analysis in order to include second order corrections $(n\leq 2)$ to 
Eq.~(\ref{lin})-(\ref{log}).  

Comparing the yellow likelihoods in Figure 
\ref{fig2} with the red ones we see that by the allowed values of $w_0$ are shifted significantly towards more 
negative values, consistent with, but broader than the kink confidence interval.  
Secondly huge values of $w_1\sim 50$ and $w_2\sim -100$ are consistent with the data. This suggests that strong dark energy dynamics is not ruled out and that higher order 
terms {\em must} be taken into account. As mentioned in the introduction, 
this comes from the fact that $w_1$ and $w_2$ are strongly degenerate in all cases and 
that there is no lower bound on $w$ at $z>1$, illustrating the huge effect of imposing the 
weak energy condition $w\geq -1$. 
 
In this case we have both a lower bound (from the weak energy 
condition) and an upper bound (from the data) on $w$. As the second order parametrisations are 
described by a parabola in their respective expansion variables, they end up being strongly 
constrained.

On the other hand, we have also considered much higher order ($n\leq 6$), and found 
that severe internal degeneracies lead to finely balanced coefficients.
Thus a Taylor expansion in $z$ becomes very unstable at high redshift, 
and even expansions in $1-a$ can hardly be called ``under control'' 
anywhere since the expansion coefficients become of order $10^3$.

\subsection{Model selection with Information Criteria and the Bayesian Evidence}

In table~\ref{tab2} we report the values of the $\chi^2$ for the best 
fit models of each parametrisation. As we can see the four-parameter Kink formula has the 
lowest $\chi^2$, followed by the scale-factor, the logarithmic and redshift parametrisation. 
Accounting for second order terms ($n\leq 2$) in the expansions provide fits even better 
than the best Kink parametrisation which is non-trivial since they have one less parameter, 
although now the allowed parameter space volume is huge. 

At this point we can ask how many parameters are actually necessary to describe the dark energy
with current SN-Ia data? Following Liddle (2004), we compute the Akaike 
information criterion (Akaike 1974)
\begin{equation}
AIC=-2\ln{\mathcal{L}}+2k
\end{equation}
and Bayesian information criterion (Schwarz 1978)
\begin{equation}
BIC=-2\ln{\mathcal{L}+k\ln{N}},
\end{equation}
where $\mathcal{L}$ is the maximum likelihood, $k$ is the number of
model parameters and $N$ is the number of data points. We also
compute the Bayesian evidence $E$ (Sivia 1996; Saini, Weller and Bridle 2004), 
both with thermodynamic integration and nested sampling (Skilling 2004).
It is worth noting that fully degenerate parameters do not contribute to the
Evidence, so that specifically the Kink model is less disfavored than the number
of parameters suggests naively. For the same reason we find that $E$ grows very slowly
when going to even higher order in the expansion-type parametrisations, although
these cases are already disfavored by Bayesian statistics. The preferred parametrisation
is $\Lambda$CDM -- it is indeed remarkable that a model with a single free parameter
fits the data so well.

\begin{table}[t]
\begin{center}
\caption{Bayesian Evidence, BIC, AIC and $\chi^2$ values of best fit for different parametrisations. 
$k$ is the total number of fitting parameters which include the dark energy 
density $\Omega_{DE}$\label{tab2}}
\begin{tabular}{crrrrr}
\tableline\tableline
model        & k & $\chi^2$ & BIC  & AIC  & $-\ln(E)$ \\ 
\tableline\tableline
$\Lambda$CDM & 1 & 177.6 & {\bf 182.7} & {\bf 179.6}  & {\bf 93} \\
w=const      & 2 & 177.6 & 187.7 & 181.6  & 96 \\
linear       & 3 & 174.5 & 189.7 & 180.5  & 99 \\
logarithmic          & 3 & 174.2 & 189.4 & 180.2  & 98 \\
scale factor & 3 & 174.0 & 189.2 & 180.0  & 98 \\
quadratic    & 4 & {\bf 172.1} & 192.3 & 180.1  & 100 \\
logarithmic II       & 4 & 172.2 & 192.4 & 180.2  & 100 \\
scale factor II    & 4 & 172.3 & 192.5 & 180.3  & 99 \\
Kink         & 5 & 172.6 & 197.9 & 182.6  & 96 \\
\tableline
\end{tabular}
\end{center}
\end{table}

\section{When did acceleration begin?}

\begin{figure}[ht]
\plotone{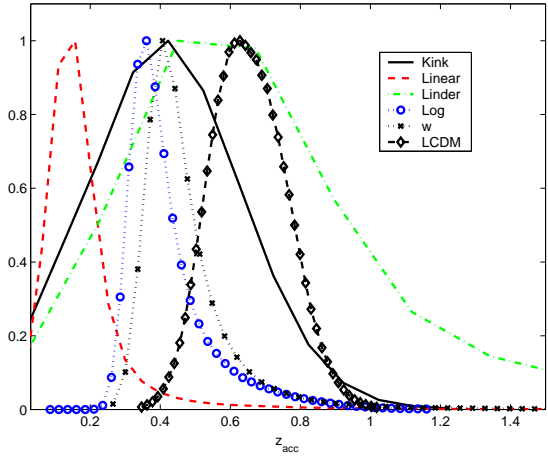}
\caption{{\bf When did acceleration begin?} The marginalised 1d likelihood for $z_{acc}$ as estimated from 
the same SN-Ia data for each of the various dark energy 
parametrisations. The wide variance casts doubt on their usefulness.
All of them peak at a lower $z_{acc}$ than in the standard 
$\Lambda$CDM, perhaps giving novel evidence 
for dark energy dynamics. Only the scale-factor expansion, eq. (\ref{linder}) 
shares significant overlap with $\Lambda$CDM, 
but that is expected since the transition is typically slow in that 
parametrisation. \label{zacc}}
\end{figure}

One of the key characteristics of dynamical dark energy is that the 
redshift at which the universe begins accelerating, $z_{acc}$, 
is characteristically different from that in $\Lambda$CDM with the same 
$\Omega_{DE}$ today. This is manifest in the CMB as a modified integrated 
Sachs-Wolfe effect \citep{Bassett,Coras3} which is degenerate with 
reionisation \citep{CHAINS}. The SN-Ia offer the possibility to break this 
degeneracy and therefore it is crucial to use a parametrisation that can 
accurately estimate $z_{acc}$ without bias.  Using a simple linear expansion 
of the deceleration parameter $q$, Riess {\em et al.} (2004) estimated 
$z_{acc} = 0.46 \pm 0.13$. Here we compare the predictions of the 
various parametrisations for $z_{acc}$. The 1d likelihoods for $z_{acc}$ are 
shown in Figure~\ref{zacc} and in table~\ref{tab3} we report the 
confidence intervals.

\begin{table}[t]
\caption{Best fit values and $1\sigma$ confidence intervals on 
$z_{acc}$ at linear order ($n\leq 1$).\label{tab3}}
\begin{center}
\begin{tabular}{crcr}
\tableline\tableline
 Model& $z_{acc}$ &Model& $z_{acc}$ \\ 
\tableline\tableline\\
$\Lambda$CDM &$0.66\pm_{0.11}^{0.11}$ &$w=$ const.      &$0.36\pm_{0.04}^{0.25}$   
\\\\ 
Linear in $z$ &$0.14\pm_{0.05}^{0.14}$&Log   &$0.38\pm_{0.08}^{0.42}$ \\\\
Scale-Factor &$0.59\pm_{0.21}^{8.91}$&Kink   &$0.45\pm_{0.44}^{0.53}$\\\\
\tableline
\end{tabular}
\end{center}
\end{table}

We have two main points. First, all of the parametrisations predict 
different best-fit values and $1\sigma$ 
error bars for $z_{acc}$, ranging from $z_{acc}=0.14$ for the redshift 
expansion to $z_{acc}=0.59$ for the scale-factor expansion (see also Dicus and 
Repko 2004).  The logarithmic, constant and Kink 
parametrisations all have similar best-fits, but the first two have overly narrow 
error bars relative to the Kink predictions. The largest error bars 
correspond to the scale-factor expansion, eq. (\ref{linder}). 

Given  the importance of accurately estimating $z_{acc}$ this variance 
forcefully argues for the need to go beyond two-dimensional parametrisations in handling future,
high-quality data. 

As a second point it is interesting that the best-fits for 
$z_{acc}$ for all the parametrisations are lower than in the $\Lambda$CDM 
model. This may be novel evidence for dark energy dynamics. We leave this issue for future work.

\section{Conclusions}
This {\em Letter} shows the limitations of standard one and two-parameter compressions of  
the infinite dimensional space of dark energy models. We have highlighted the 
dangers in using constraints  derived using these parametrisations, 
particularly regarding the possibility of rapid evolution in the dark energy, 
which none of the standard compressions can follow, and in defining allowed regions of
parameter space which are completely wrong in the case where 
the weak energy condition ($w>-1$) is not imposed. 

Rapid evolution provides a superlative fit to 
current SN-Ia data (as measured by $\chi^2$), despite claims to the contrary in the 
literature which were based on two-parameter compressions. Indeed, all of the two-parameter
expansions we studied wrongly rule out such rapid evolution at $2\sigma$ or more. 
This is extremely damning evidence, especially since these compressions are typically used in the 
planning for the next-generation experiments which will provide data of significantly higher quality. 
In addition, the standard parametrisations  also miss the fact that $w$ has no 
lower-bound at $z>1$ if the weak energy condition is not imposed, artifically cutting-out vast swaths 
of parameter space due to their innate limitations.

Further problems occur in estimating the redshift at which the universe began 
accelerating, $z_{acc}$. There is a  nearly 
300\% variation in the best-fit for $z_{acc}$, depending on parametrisation.
Interestingly, all the tested parametrisations gave best-fits for $z_{acc}$ below that of 
$\Lambda$CDM, providing unusual cross-parametrisation evidence 
for dark energy dynamics. Nevertheless, use of the Bayesian information criteria 
for model selection prefers the cosmological constant over the other models which 
remains the model to beat. 

The severe inadequacy of the standard two-parameter expansions lead us to 
consider higher-order terms ($n\geq 2$) with one of more extra parameters, e.g. $w_2$.
While this brings the rapid evolution models within the allowed region of parameter 
space it leads to severe degeneracies (see figure \ref{fig2}) which may make the 
parametrisations impotent for constraining the space of theoretical dark energy models, particularly
when $w<-1$.  

For completeness we have also compared four standard parametrisations against a test-bed 
of quintessence models (Appendix) and found that while the kink and scale-factor expansion are
excellent, the expansions in $z$ and $\log(1+z)$ can lead to large errors for $z \geq 1$. 

We conclude that confidence intervals inferred from standard 
two-parameter expansions often do not deserve that name and are typically untrustworthy, 
even with current data. The wealth and quality of dark energy data we will
aquire over the next decade will demand significantly better performance. 

\acknowledgments
It is a pleasure to thank Rob Caldwell, Rob Crittenden, Ruth Durrer, Dragan Huterer, 
Eric Linder, Andrew Liddle, Roy Maartens, Takahiro Tanaka and Jochen Weller for discussions or comments. 
P.S.C. is grateful to the  Michigan Center for Theoretical Physics for 
hospitality during the ``Dark Side of the 
Universe'' workshop where part of this work was done.
P.S.C. is supported by Columbia Academic Quality Fund. M. K. is
supported by PPARC while BB is supported by a Royal Society/JSPS 
fellowship.

\appendix
%\twocolumn
%\section*{Appendix}
In this appendix we address the problem of how accurately the various 
parametrisations  listed earlier do in capturing the dynamics of standard quintessence 
models.  We proceed by considering a test-bed of four popular quintessence 
models, which provide a wide range of different  behaviours for $w(z)$, with 
smooth, slow and rapid transitions. 
%%%%%%%%%%%%%%%%%%%%%%%%%%%%%%%%
\begin{figure}[ht]
\plotone{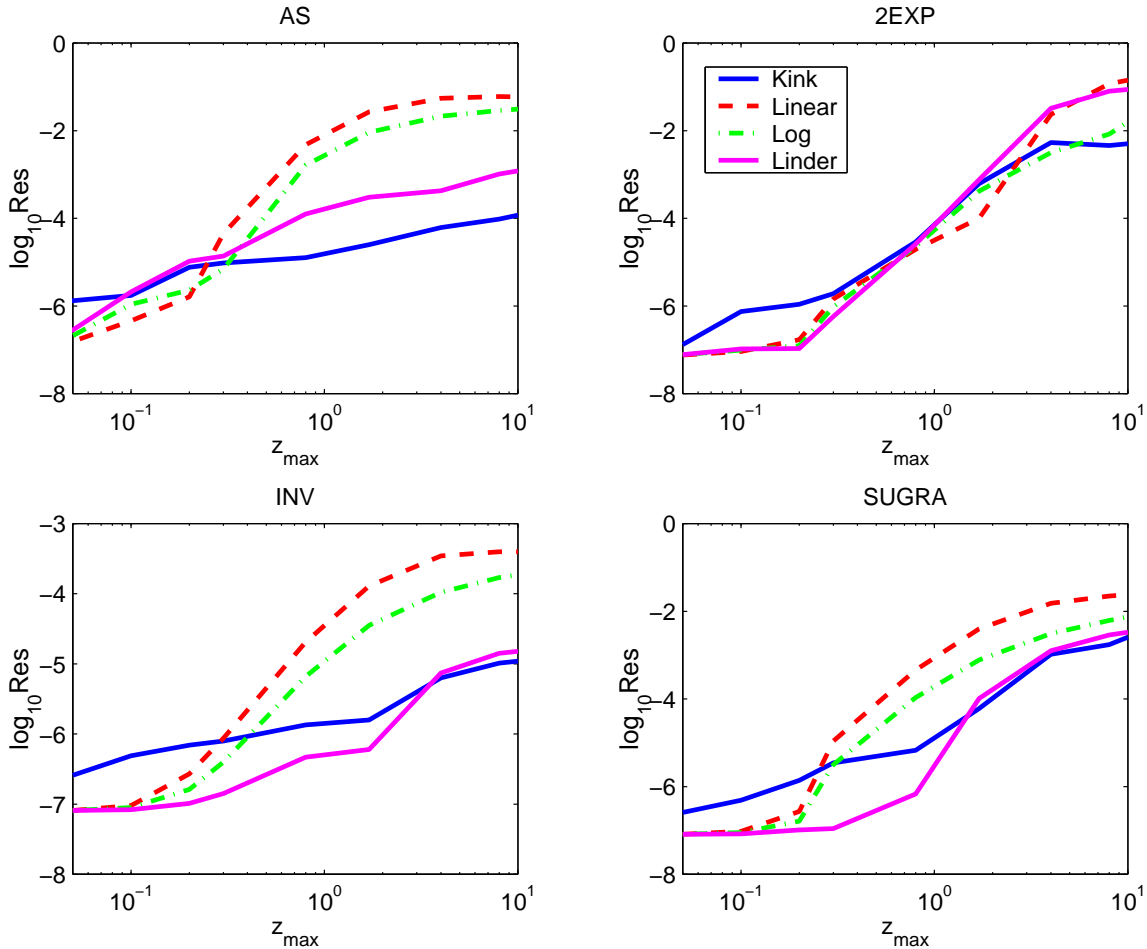}
\caption{Plot showing the mean {\em quadratic} error of the best-fit, 
eq. (\ref{res}), to the different quintessence models as function of $z_{max}$ for each 
parametrisation for $n\leq 1$. At $z\sim 1$ the expansion to 
first order in $z$ (`redshift') and $\log(1+z)$ (`logarithmic') typically 
show errors around 10\% while the kink and scale-factor parametrisations have 
errors typically at the 1\% level at $z \sim 1$.  \label{fig1}}
\end{figure}
%%%%%%%%%%%%%%%%%%%%%%%%%%%%%%%%
Our test-bed suite of models is:
the Albrecht-Skordis model (Albrecht and Skordis 2001), 
\begin{equation}
V(Q)=M^4 e^{-\alpha Q}[(Q-A)^2+B]^{s/2},
\end{equation}
with $\alpha=5$, $A=54.5$, $B=0.01$ and $s=5$; 
the Two Exponential potential (Barreiro, Copeland and Nunes 2000),
\begin{equation}
V(Q)=M^4(e^{-\alpha Q}+e^{-\beta Q}),
\end{equation} 
with slopes $\alpha=-9$ and $\beta=2$; 
the inverse power-law potential (Ratra and Peebles 1988),
\begin{equation}
V(Q)=M^{4+\alpha}/Q^{\alpha},
\end{equation} 
with $\alpha=0.6$; 
the Supergravity (SUGRA) inspired model (Brax and Martin 1999), 
\begin{equation}
V(Q)=M^4 e^{-Q^2/2}/Q^{\alpha},
\end{equation} 
with $\alpha=6$.
For the kink and each two-parameter expansion ($n\leq 1$) we compute the best-fit (using an MCMC search) to the different
quintessence models in the redshift range $0<z<z_{max}$ and plot the average residual {\em quadratic 
errors} as a function of $z_{max}$, as shown in Figure \ref{fig1}, where
\begin{equation}
Res=\frac{1}{N_z} \sum_{i=1}^{N_z} (w_Q(z_i)-w_{par}^{best}(z_i))^2,
\label{res}
\end{equation}
with $N_z$ the total number of equally spaced bins in the range 
$[0,z_{max}]$, and we have chosen $N_z=1000$. Comparing the residuals indicates how well 
each formula reproduces a given  quintessence model as function of the redshift interval.  For 
$z_{max}\ge 0.3$, the Kink and scale-factor formulae provide best-fits which are at 
least one order of magnitude better than the redshift and logarithmic expansions, 
except perhaps in the Two-Exponential case. 

Only for $z_{max}<0.3$ are the residuals of
the redshift and logarithmic formulas of the same order as
those of the other parametrisations. This is easy to understand from 
Fig. (\ref{fig6}). The best-fit solution is almost perfectly 
linear for $z < 0.1$ but then suddenly 
deviates. A linear fit will be excellent until $z \sim 0.1$, after which it rapidly becomes 
bad.

We do not  consider the case of constant $w$ which can at best fit the average value of $w(z)$ and whose
errors are expected to be the worst.

\clearpage


\begin{thebibliography}{}
\bibitem[]{aka} Akaike, H., 1974, IEEE Trans. Auto. Control, 19, 716.
\bibitem[Alam et al. 2003b]{Sahni04a} Alam, U., Sahni, V, Saini, T.D. 
and Starobinsky, A.A. 2003, astro-ph/0311364.
\bibitem[]{al} Albrecht, A. and Skordis, C. 2001, Phys. Rev. 
D\textbf{64}, 023514, astro-ph/9908085.
\bibitem[]{barre} Barreiro, T., Copeland, E.J. and Nunes, N.J. 2000,
Phys. Rev. D\textbf{61}, 127301, astro-ph/9910214.
\bibitem[Bassett {\em et al.} 2002]{Bassett} Bassett, B.A., Kunz, M., 
Silk, J. and Ungarelli, C. 2002, \mnras \textbf{336}, 1217, astro-ph/0203383.
\bibitem[Bassett {\em et al.} 2003]{condens} Bassett, A. M., Kunz, M., 
Parkinson, D. and Ungarelli, C., 2003, Phys. Rev. D\textbf{68}, 043504, astro-ph/0211303.
\bibitem[]{BK04} Bassett, B. A. and Kunz, M., 2004,  Phys. Rev. D\textbf{69}, 101305, astro-ph/0312443.
\bibitem[]{Brax} Brax, P. and Martin, J. 1999, Phys. Lett. 
B\textbf{468}, 40, astro-ph/9905040.
\bibitem[]{CD} Caldwell R.~R. and Doran, M., 2004, Phys. Rev. D \textbf{69}, 103517 
\bibitem[]{che} Chevallier, M. and, Polarski, D. 2001, 
Int. J. Mod. Phys. D\textbf{10}, 213, gr-qc/0009008.
\bibitem[Corasaniti and Copeland 2002]{Coras2} Corasaniti, P.S. and 
Copeland, E.J. 2002, Phys. Rev. D\textbf{67}, 063521, astro-ph/0205544.
\bibitem[Corasaniti {\em et al.} 2003]{Coras3} Corasaniti, P.S., 
Bassett, B.A.,
Copeland, E.J. and Ungarelli, C. 2003, Phys. Rev. Lett. \textbf{90},
091393, astro-ph/0210209.
\bibitem[Corasaniti {\em et al.} 2004]{CHAINS} Corasaniti, P.S., Kunz, 
M., Parkinson, D.,
Copeland, E.J. and Bassett, B.A., 2004, astro-ph/0406608.
\bibitem[]{dicus} Dicus, D.A. and Repko, W.W. 2004, astro-ph/0407094.
\bibitem[Efstathiou 1999]{Efstathiou99} Efstathiou, G. 1999, 
\mnras \textbf{303}, L47, astro-ph/9812226.
\bibitem[Feng {\em et al.} 2004]{Feng} Feng, B., Wang, X.-L. and Zhang, 
X.-M. 2004, astro-ph/0404224.
\bibitem[Gong 2004]{Gong} Gong, Y. 2004, astro-ph/0405446.
\bibitem[Hu 2004]{hu04} Hu, W. and Jain, B., (2003) astro-ph/0312395.
\bibitem[Huterer and Turner 2001]{Huterer00} Huterer, D. and Turner, 
M.S. 2001, Phys.Rev. D\textbf{64}, astro-ph/0012510.
\bibitem[Huterer and Starkmann 2003]{Huterer02} Huterer, D. and 
Starkman, M.S. 
2003, Phys. Rev. Lett. \textbf{90}, 031301, astro-ph/0207517. 
\bibitem[]{co} Huterer, D. and Cooray, A.R. 2004, astro-ph/0404062.
\bibitem[Jassal {\em et al.} 2004]{Jassal} Jassal, H.K., Bagla, J.S. 
and Padmanabhan, T. 2004, astro-ph/0404378.
\bibitem[Jonsson {\em et al.} 2004]{Jonsson} Jonsson, J, Goobar, A., 
Amanullah, R.
and Bergstrom, L. 2004, astro-ph/0404468.
\bibitem[]{ku} Kunz, M., Corasaniti, P.S., Parkinson, D. and Copeland, 
J.C. 2003, astro-ph/0307346.
\bibitem[]{kb} Kunz, M. and Bassett, B. A., 2004, astro-ph/0406013.
\bibitem[Liddle 2004]{Liddle04} Liddle, A.R. 2004, astro-ph/0401198.
\bibitem[Linder 2003]{Linder03} Linder, E.V. 2003, Phys. Rev. Lett. 
\textbf{90}, 91301, astro-ph/0402503.
\bibitem[Linder 2004]{Linder04} Linder, E.V. 2004, astro-ph/0402503.
\bibitem[Maor {\em et al.} 2002]{ma} Maor, I., Brustein, McMahon, J. 
and Steinhardt, P.J.
2002, Phys. Rev. D\textbf{65}, 123003, astro-ph/0112526.
\bibitem[Melchiorri {\em et al.} 2003]{alm} Melchiorri, A. {\em et al.}, 
2003, Phys.\ Rev.\ D {\bf 68}, 043509.
\bibitem[]{ra} Ratra, B. and Peebles, P.J.E. 1988, Phys. Rev. 
D\textbf{37}, 3406. 
\bibitem[Riess {\em et al.} 2004]{Riess} Riess, A. et al. 2004, 
astro-ph/0402512.
\bibitem[]{swb} Saini, T.D., Weller, J. and Bridle, S.~L., 2004,
Mon.\ Not.\ Roy.\ Astron.\ Soc.\  {\bf 348}, 603 
\bibitem[]{scw} Schwarz, G. 1978, Annals of Statistics, 5, 461.
\bibitem[]{sivia} Sivia, D.~S.~1996, Data Analysis: A Bayesian
Tutorial (New York: Oxford Univ. Press)
\bibitem[]{skilling} Skilling, J. 2004, 
{\tt http://www.inference.phy.cam.ac.uk/bayesys/}
\bibitem[Virey {\em et al.} 2004]{Virey} J.~M.~Virey, P.~Taxil, 
A.~Tilquin, A.~Ealet, D.~Fouchez and C.~Tao, arXiv:astro-ph/0403285.
\bibitem[Wang and Freese 2004]{Wang} Wang, Y., F. and Freese, K. 2004, 
astro-ph/0402208.
\bibitem[Wang and Tegmark 2004]{WT} Wang, Y., and Tegmark, M., 2004, Phys. Rev. Lett., \textbf{92}, 241302
\bibitem[Weller and Albrecht 2002]{Weller01} Weller, J. and Albrecht, 
A. 2002,
Phys. Rev.D\textbf{65}, astro-ph/0106079. 
\bibitem[Wetterich 2004]{Wetterich} Wetterich, C. 2004, astro-ph/0403289.
\end{thebibliography}
\end{document}